\documentclass[final,3p,times,twocolumn]{elsarticle}
\usepackage{graphicx}
\usepackage{amssymb}
\journal{Physics Letters B}

\begin{document}

\begin{frontmatter}

\title{Implications of the ABC Resonance Structure on Elastic Neutron-Proton
  Scattering}   

\author[PIT,KEP]{Annette Pricking}
\author[PIT,KEP]{M. Bashkanov}
\author[PIT,KEP]{H.~Clement\corref{Phys}
%\ead{clement@pit.physik.uni-tuebingen.de}
}
\cortext[Phys]{corresponding author: H. Clement} 
\ead{clement@pit.physik.uni-tuebingen.de}

\address[PIT]{Physikalisches Institut der Universit\"at T\"ubingen, Germany}
\address[KEP]{Kepler Center for Astro and Particle Physics,
 University of T\"ubingen, Auf~der~Morgenstelle~14, D-72076 T\"ubingen,
 Germany}

\begin{abstract}
In recent WASA-at-COSY measurements of the basic double-pionic fusion
reactions $pn \to d\pi^0\pi^0$  and $pn \to d\pi^+\pi^-$ 
%a strict correlation between the ABC effect and 
a narrow resonance structure with $I(J^P) = 0(3^+)$ in the total cross section
has been found. If this constitutes a $s$-channel resonance 
%with definite spin-parity 
in the $pn$ system, then it should cause distinctive consequences
in $pn$ scattering. The magnitude of the decay width into the $pn$ channel is
estimated and the expected resonance effects in integral and differential $pn$
scattering observables are presented. The inclusion of the resonance improves
the description of total cross section data. For the analyzing power a
characteristic energy dependence is predicted, which should allow a crucial 
experimental check of the resonance hypothesis. 
\end{abstract}


\end{frontmatter}

\section {Introduction}

The so-called ABC-effect, which constitutes a peculiar low-mass enhancement in
the invariant mass of an isoscalar pion pair produced in a double-pionic
fusion reaction, has been a puzzle all the time since its first discovery
fifty years ago by Abashian, Booth and Crowe \cite{abc}. 
Recent WASA-at-COSY experiments \cite{adl,isofus} on the basic double-pionic
fusion to deuterium established a tight correlation between the
appearance of the ABC effect and a narrow Lorentzian energy dependence with
mass m = 2.37 GeV and width $\Gamma$ = 70 MeV in the
integral cross sections of the reactions $pn \to d\pi^0\pi^0$ and $pn \to
d\pi^+\pi^-$, isoscalar part. The differential distributions are consistent
with a $I(J^P) =  0(3^+)$ assignment to this resonance-like structure. In
addition the experimental Dalitz plots point to a $\Delta\Delta$
excitation in the intermediate state. Hence we consider the following reaction
scenario for the interpretation of the data:

\begin{equation}
pn \to R \to \Delta\Delta \to (NN\pi\pi)_{I=0},
\end{equation}

where R denotes a $s$-channel resonance in $pn$ and $\Delta\Delta$ systems. By
this scenario we explictly neglect a possible direct decay $R \to
NN\pi$. Note that an intermediate $N\Delta$ configuration is excluded by
isospin. 

In this paper we consider the possible decay channels of such a resonance in the
scenario of eq. (1). In particular we estimate the partial decay width into
the elastic $pn$ channel and calculate the effect of such a resonance onto the
$pn$ scattering observables.

\section {Decay channels and widths}

The cross section of the  isoscalar two-body resonance process $pn \to R \to
\Delta\Delta$ is given by
\begin{eqnarray}
\sigma_{pn \to \Delta\Delta} = \frac {4 \pi} {k_i^2}  \frac {2J+1} {(2s_p +
  1) (2s_n + 1)}  
\frac {m_R^2 \Gamma_i
\Gamma_f} {(s-m_R^2)^2 +m_R^2 \Gamma^2},
\end{eqnarray}
where $k_i$ denotes the initial center-of-mass momentum.

With $J$ = 3 and $s_p = s_n$ = 1/2 the peak cross section at $\sqrt s =
m_R$ = 2.37 GeV ($k_i$ = 0.72 GeV/c) is then 
\begin{equation}
\sigma_ {pn \to \Delta\Delta}(peak) = \sigma_0 \frac  {\Gamma_i \Gamma_f }
{\Gamma^2}
\end{equation}
with 
\begin{equation}
\sigma_0 = 16.4~mb ~~(unitarity~limit).
\end{equation}

Since we also have

\begin{equation}
\Gamma = \Gamma_i + \Gamma_f,
\end{equation}
we get from (3) and (5):
\begin{equation}
\Gamma_i = \Gamma ( \frac 1 2 \pm \sqrt { \frac 1 4 - \frac
  {\sigma_{pn \to \Delta\Delta}(peak)} {\sigma_0 }}).
\end{equation}

To estimate $\sigma_{pn \to \Delta\Delta}(peak)$ consider the total cross
sections of all channels, where the isoscalar $\Delta\Delta$ system can decay
into:  
\begin {itemize}
\item (i) $d\pi^0\pi^0$ and $d\pi^+\pi^-$: \\
  Due to isospin rules we expect
\begin{equation}
\sigma_{d\pi^+\pi^-}(I=0)~=~2~\sigma_{d\pi^0\pi^0}, 
\end{equation}
however, due to the isospin violation in the pion mass, the available phase
space is somewhat smaller for charged pion production than for the production
of the lighter neutral pions. In Ref. \cite{isofus} it has been shown that
this results in a resonance cross section, which is lower by about 20 $\%$ in
case of the $d\pi^+\pi^-$ channel. Hence we have
\begin{equation}
\sigma_a := \sigma_{d\pi^+\pi^-} + \sigma_{d\pi^0\pi^0} \approx 2.6~
\sigma_{d\pi^0\pi^0}.
\end{equation}

%where we neglect the small isospin violation effect due to different masses of
%charged and neutral pions -- see discussion in Ref. \cite{isofus}.
The peak cross section of the $pn \to d\pi^0\pi^0$ reaction at  $\sqrt s$ = 2.37
GeV  has been measured to be 0.27 mb \cite{isofus}. This includes the
contributions of the 
%of the interference of the resonance with background amplitudes due to the
$t$-channel $\Delta\Delta$ and Roper excitations. Accounting for this
background effect the pure resonance cross section in this channel amounts to
about 0.24 mb,  
{\it i.e.}:
\begin{equation}
\sigma_a \approx 0.6~mb.
\end{equation}
\item (ii) $np\pi^0\pi^0$, $np\pi^+\pi^-$ and $pp\pi^0\pi^-$
  - only I = 0 part: \\
 
In a recent paper \cite{WF} F\"aldt and Wilkin present an estimate of the
resonance cross section in the $pn \to pn\pi^0\pi^0$ reaction. According to
their calculation based on final state interaction theory the expected peak
cross section in the deuteron breakup channel $pn\pi^0\pi^0$ is about 85$\%$
that of the non-breakup channel $d\pi^0\pi^0$, {\it i.e.} about 0.2 mb. Very
recently also Albaladejo and Oset \cite{oset} estimated the expected resonance
cross sections  in $pn \to pn\pi^0\pi^0$ and $pn \to pn\pi^+\pi^-$ using a more
elaborate theoretical procedure. Their result for the $pn \to pn\pi^0\pi^0$
channel is compatible with that from Ref. \cite{WF}.

Next we consider the $pp\pi^0\pi^-$ channel. Though both the $pp$ pair and the
$\pi^0\pi^-$ pair are isovector pairs, they
may couple to $I = 0$ in total. Hence the isoscalar resonance may also
decay into the isoscalar part of the $pp\pi^0\pi^-$ channel. In fact, 
the decay of the resonance into the $pp\pi^0\pi^-$ channel proceeds via the
same intermediate $\Delta^+\Delta^0$ system as the $d\pi^0\pi^0$ channel
does. From isospin coupling we expect that the resonance decay into the
$pp\pi^0\pi^-$ system should be half that into the $np\pi^0\pi^0$ system. And
since from the estimates in Ref. \cite{WF} we expect the resonance effect in
the $np\pi^0\pi^0$ system to be about 0.20 mb, we estimate the peak resonance
contribution in the $pp\pi^0\pi^-$ system to be in the order of 0.1 mb. In
fact, a recent measurement \cite{ts} of this channel by WASA-at-COSY is in
agreement with such a resonance contribution in the total cross section
at $\sqrt s$ = 2.37 GeV.   

The resonance effect in the isoscalar part of the $np\pi^+\pi^-$ channel is
composed of the configurations, where either both $np$ and $\pi^+\pi^-$ pairs
couple each to $I~=~0$ or both pairs each to $I~=~1$. The latter case provides
the same situation as the $pp\pi^0\pi^-$ channel. Hence we have

\begin{eqnarray}
\sigma_{np\pi^+\pi^-}(I = 0) \approx 2 \sigma_{np\pi^0\pi^0} +
\sigma_{pp\pi^0\pi^-} \\ \nonumber
\sigma_b := \sigma_{np\pi^+\pi^-} + \sigma_{np\pi^0\pi^0} +
\sigma_{pp\pi^0\pi^-} \\
\approx 0.5~mb + 0.2~mb + 0.1~mb \\
\approx 0.8~mb. \nonumber
\end{eqnarray}

We note that our estimate for the resonant $pn \to pn\pi^+\pi^-$ cross section
is in agreement with that of Ref. \cite{oset}.  

\item (iii) $pp\pi^-$ and $pn\pi^0$ (I=0 part):\\
    The isoscalar part of single-pion production is not well
    known. Recent work \cite {sar04,sar10} suggests a maximum isoscalar
    cross section at $\sqrt s$ = 2.30 GeV with an indication of some steep
    decline thereafter. 
    At our resonance energy there are no data at all. Independent of
    this it is very hard to construct a process, where the intermediate
    $\Delta\Delta$ system decays by emission of a single pion only. In such a
    case 
    one of the $\Delta$ excitations must be de-excited by pion exchange with the
    other $\Delta$. However, the formation of an intermediate $N\Delta$ state
    is isospin forbidden -- as already mentioned in the introduction. Also, the
    condition $J^P = 3^+$ is very hard to fulfill in 
    such a scenario. 
%Since the single pion is emitted in relative $p$-wave
%    from the de-exciting $\Delta$, the two nucleons in the final states would
%    need to have large relative orbital angular momentum, {\it i.e.} would
%    need to be at least in a $^1D_2$ state - again a very unlikely
%    scenario. 
    Hence we conclude that any decay of the resonance R into these
    single-pion channels must be small compared to the favored
    decays into the two-pion channels.
\end {itemize}

Altogether we get as an estimate 
\begin{equation}
\sigma_{pn \to \Delta\Delta}(peak) = \sigma_a + \sigma_b  \approx 1.4(1)~mb.%%%!!!!!
\end{equation}
Putting this into eq. (6) and selecting the minus sign before the root we obtain
\begin{equation}
\Gamma_i = 7(1) MeV  ~~~~~~~~for~\Gamma = 70~MeV,
\end{equation}
which in turn corresponds to a resonance cross section in the elastic pn
channel of only
\begin{equation}
\sigma_{pn \to pn} \approx 0.16~mb,
\end{equation}
if the resonance would contribute only incoherently.

From the peak cross sections given under (i) and (ii) as well as from eqs. (3)
- (13) we may readily calculate the branching ratios BR := $\Gamma_j /
\Gamma$ for the decay of the resonance into the channels j. The results are
listed in Table 1.

\begin{table}
\caption{Branching ratios of the $d^*$ resonance into its decay channels based
on eqs. (3) and (12) and the peak cross sections given under (i) and (ii).}  
\begin{tabular}{lllll} 
\hline

% & ~~~~~~~~~~~~~~~~~~& $\sigma_{tot}$ [$\mu$b]~~~~~~~~~~\\

 & decay channel&branching ratio&remarks&\\ 

\hline

& $np$ & 10 $\%$ & predicted &  \\
& $d\pi^0\pi^0$ & 15 $\%$& measured &  \\
& $d\pi^+\pi^-$ & 25 $\%$ & measured &  \\
& $pp\pi^0\pi^-$ & 7 $\%$ & measured &  \\ 
& $np\pi^+\pi^-$ & 31 $\%$ & predicted &  \\ 
& $np\pi^0\pi^0$ & 12 $\%$ & predicted &  \\ 
\hline
 \end{tabular}\\
\end{table}

The value obtained for $\Gamma_i$ appears to be quite reasonable. It is
somewhat smaller than  
the quark-model predictions  of Ping et al. \cite {ping} (see
their Table V) where they quote $\Gamma_i$ = 9 - 17 MeV. In this table they
also quote a value of $\Gamma_i \leq$ 18 MeV to be consistent with the SAID phase
shift analysis SP07 \cite{said}. An upper limit for $\Gamma_i$ may be
directly derived also from Table~2, where SAID cross sections are quoted for
selected partial waves. Since $J^P = 3^+$, the initial partial waves for the
formation of the resonance R are the $^3D_3$ and/or $^3G_3$ $pn$ partial
waves. The $^3D_3$ total elastic cross section at $T_p$ = 1.2 GeV is 1.46
mb. Taking this as an upper limit for the elastic resonance cross section we
obtain as an upper limit for the elastic decay width $\Gamma_i \leq$ 20
MeV. In case of a resonance excitation purely by the $^3G_3$ partial wave the
total elastic cross section given by SAID for this partial wave is already
exhausted by $\Gamma_i$ = 9 MeV. We note, however that we discuss here two
extreme situations. Actually, $^3D_3$ and $^3G_3$ are J-coupled partial waves
allowing for a mixing of both components. {\it I.e.}, the true solution may be
in-between the two extreme cases, which we discuss in this paper for simplicity. 

We note in passing that the other solution of eq. (6) -- the one with the
+sign -- leads to $\Gamma_i$ = 62 MeV implying that the resonance would be
predominantly elastic. i.e. mainly decaying into the elastic channel and only
weakly decaying into the pion-production channels. This solution is at obvious
variance with SAID.

%We also note that since the $pn$ threshold is far away from the resonance
%pole, $\Gamma_i$ can be approximated to be energy-independent over the region
%of the resonance. 

Before we continue to discuss the consequences of the resonance hypothesis for
the $pn$ scattering observables, we shortly want to discuss the situation for
the case that the spin-parity of the resonance would have been $J^P = 1^+$.
As discussed in Ref. \cite{adl} a $\Delta\Delta$ system  in relative $s$-wave
in the intermediate state could in principle have $J^P = 1^+$ or $3^+$.
In the  $J^P = 1^+$ case we would get the unitarity limit $\sigma_0$ = 7.0 mb
and using the estimate of F\"aldt and Wilkin \cite{WF}
$\sigma_{\Delta\Delta}(peak) / \sigma_0$ = 0.31. According to eq. (6) this
leads, however, to an imaginary part for the partial width $\Gamma_i$ in the
$pn$ channel. To avoid this imaginary part necessitates
$\sigma_{\Delta\Delta}(peak) \leq$ 1.75 mb. This in turn means $\sigma_b \leq$
0.85 mb, which is at variance with the estimates of F\"aldt and Wilkin
\cite{WF}. Taking this limiting case would result in  
$\Gamma_i = \Gamma / 2$ = 35 MeV and $\sigma_{pn}$ = 1.75 mb. As already
demonstrated by F\"aldt and Wilkin \cite{WF} the estimated cross section for J
= 1 exceeds the sum of the SAID inelastic cross sections in the $^3S_1$ and 
$^3D_1$ partial waves. 

%As given in Tab. 2 we have at $T_p$ = 1.2
%GeV the following situation for the $NN$ partial waves under discussion. 
\begin{table}
\caption{Total, elastic and reaction cross sections for selected isoscalar
  $pn$ partial waves at $T_p$ = 1.2 GeV according to SAID \cite{said}.}  
\begin{tabular}{lllll} 
\hline

% & ~~~~~~~~~~~~~~~~~~& $\sigma_{tot}$ [$\mu$b]~~~~~~~~~~\\

 & partial wave&$\sigma_{tot}$[mb]&$\sigma_{tot}^{el}$[mb]&$\sigma_{tot}^{reac}$[mb]\\ 

\hline

& $^3S_1$ & 7.05 & 6.23 & 0.82 \\

& $^3D_1$ & 4.51 & 3.17 & 1.34 \\

& $^3D_3$ & 5.95 & 1.46 & 4.49 \\
& $^3G_3$ & 1.30 & 0.25 & 1.05 \\ 

\hline
 \end{tabular}\\
\end{table}
%The $^3D_3$ partial wave provides by far the largest reaction
%cross section. As mentioned above $^3S_1$ and $^3D_1$ would only contribute to
%a $J^P = 1^+$ resonance. For the $J^P = 3^+$ case $^3D_3$ and $^3G_3$ can
%contribute, however, the $^3G_3$ partial wave is much weaker, both elastically
%and inelastically than the $^3D_3$ partial wave. Since the  $^3G_3$ reaction
%cross section is much smaller than the resonance cross sections in $NN\pi$ and
%$NN\pi\pi$ channels, this partial wave can not be a
%major source for the resonance formation in the incident $pn$ channel.   
%For a more quantitative estimate of the contribution  of the $^3G_3$  partial
%wave, which would be a spin-flip contribution to the resonance formation, one
%would need a microscopic understanding of this transition.

In general the decay widths of a resonance are momentum dependent. This is
important, if we consider the resonance not only at its resonance mass -- as
done above -- but also over a wider range of energies, as we will do now in
the following. The momentum dependence is particularly significant for the
numerator of the resonance amplitude, where the  
elastic decay width enters linearly and is highly momentum dependent due to
the $D$- and $G$-wave character, respectively of the relevant partial
waves. Following Ref. \cite{teis} we parameterize the elastic width due to the
resonance excitation in the $^3$L$_3$ partial wave as follows:
\begin{equation}
\Gamma_i(q) = \Gamma_i (\frac q q_R)^{2L+1} (\frac{q_R^2 + \delta^2}{q^2 +
  \delta^2})^{L+1}, 
\end{equation}
where $q$ and $q_R$ are the nucleon three-momenta in the rest-frame of the
resonance at energies $\sqrt s$ and $m_R$, respectively. For the cutoff
parameter we use $\delta$ = 0.5 GeV/c$^2$.

In the exit channel the resonance decays into the $\Delta\Delta$ system with a
relative s-wave between the two $\Delta$s --- as observed in the $\Delta$
angular distribution (Fig. 5 in Ref. \cite{adl}). Therefore we have 

\begin{equation}
\Gamma_{\Delta\Delta} = g_{\Delta\Delta}^2 q_{\Delta\Delta}
F(q_{\Delta\Delta})^2
\end{equation}
where a monopole form-factor 
\begin{equation}
F(q_{\Delta\Delta}) = \frac {\Lambda^2} {\Lambda^2 + q_{\Delta\Delta}^2/4}
\end{equation} 

is introduced, in order to account for the ABC effect (see Refs. \cite{adl,MB}.
The cutoff parameter $\Lambda$ is adjusted for best reproduction of the ABC
effect (low-mass enhancement) in the $M_{\pi\pi}$ spectrum. Since
$q_{\Delta\Delta}$ = $q_{\pi\pi}$, when neglecting the Fermi motion of the
nucleons, this form-factor is reflected directly in the $M_{\pi\pi}$ spectrum
and causes there the ABC effect by suppression of the high-mass region. Fitting
the cutoff parameter $\Lambda$ of this monopole form-factor to
the data in the  $M_{\pi\pi}$ spectrum results \cite{adl} in
\begin{equation}
\Lambda \approx  0.16~GeV/c
\end{equation}
corresponding to a length scale of $r = \frac {\hbar \sqrt 6} {\Lambda}
\approx$ 2 fm. 
%({\bf Interpretation ??}) 

%Note that the form-factor works primarily on the $M_{\pi\pi}$ spectrum as
%explained above. But this is not the only location to sense the
%form-factor. Indirectly it also influences other observables - such as
%the $M_{d\pi}$ distribution (as we show in the appendix) and also angular
%distributions, since it causes the reaction's intensity to concentrate in a
%small kinematic range around small $\pi\pi$-invariant masses, see Dalitz plots
%in Fig. 8. In all those cases the effect is small, but significant -- and
%it always improves the description of the data.

The total width of the resonance is then given by 
\begin{eqnarray}
&&\Gamma_R(s) = \Gamma_i + \sum \Gamma_f = \Gamma_i(q) +  \gamma_R\\ 
&&\int dm_1^2 dm_2^2
q_{\Delta\Delta} F(q_{\Delta\Delta})^2 | D_{\Delta_1}(m_1^2)
D_{\Delta_2}(m_2^2)|^2 \nonumber,
\end{eqnarray}
where the integral runs over all possible $q_{\Delta\Delta}$ and $N\pi$-invariant
mass-squared $m_1^2$ ($m_2^2$) forming the systems $\Delta_1$ and
$\Delta_2$, respectively \cite{CH}. 

%Here $\Gamma_0$ stands for the sum of decay widths into $NN$ and $NN\pi$
%channels. Since the threshold of these channels is far away, these widths may
%be approximated to be momentum independent over the range of the
%resonance. From the study of possible resonance effects in the cross
%sections of these channels (see next section)  a value of $\Gamma_0 \approx$
%30 MeV can be estimated. 

The second term in eq. (18) denotes the decays of the resonance via the
intermediate $\Delta\Delta$ system. The quantity $\gamma_R$  contains the
coupling constant $g_{\Delta\Delta}$ and other constants and is fitted to
yield a total width of $\Gamma_R(s = m_R^2)$ = 70 MeV. 

%In the $pn \to d\pi^0\pi^0$ channel, where the resonance effect dominates an
%energy range of about 200 MeV in $\sqrt s$, the effect of the momentum
%dependence of the decay widths on the total cross section is minor as
%demonstrated in Fig.1, bottom. Still, on the low-energy end it gets sizable.
%In $pn$ scattering we expect large interference effects of the resonance
%amplitude with the non-resonant amplitudes in the $J^P = 3^+$ partial
%waves. Hence we have to consider there larger energy ranges ranges, where the
%momentum dependence gets more important.

%The relative size of the two terms in (2.11) is not very critical for
%the energy dependence of the total cross section. This is
%demonstrated in Fig.2, where we compare the cases $\Gamma_0$ = 0, 35 and 70
%MeV for $\Gamma(s = m_R^2)$ = 70 MeV.

%\begin{figure} 
%\centering
%\includegraphics[width=0.99\columnwidth]{CH_DD_width1.eps}\\
%\includegraphics[width=0.32\columnwidth]{CF_flat.eps}
%\includegraphics[width=0.32\columnwidth]{CF_HCl_Mod.eps}
%\includegraphics[width=0.99\columnwidth]{CF_Mod.eps}
%\caption{\small {\bf Top:} energy dependence of the total resonance width
%  $\Gamma (s)$ according to eq. (18) with  $\Gamma (\sqrt s$~=~2.37~GeV) = 70
%  MeV. {\bf Bottom:} Total cross section of the $pn \to d\pi^0\pi^0$ reaction
%  calculated with either an momentum 
%  independent total width of $\Gamma (s)$ = 70 MeV (dotted) or with the
%  momentum dependent width (solid) shown in the top figure. The plotted
%  symbols denote the data from Ref. \cite{adl}.
%}
%\label{fig1}
%\end{figure}

\section{Resonance amplitude in the $pn$ channel}

Knowing now the partial decay width of the resonance R into the elastic $pn$
channel we can calculate the resonance effect in this channel by adding the
resonance amplitude to the corresponding partial wave amplitude of the energy
dependent SAID solution.

The scattering amplitude is given by the T-matrix elements for the $(l,j)$th
partial wave, which are connected to those of the S-matrix by 
\begin{equation}
%T_{lj}^{SAID} = \frac{S_{lj}^{SAID} - 1} {2i} 
T_{lj} = \frac{S_{lj} - 1} {2i}. 
\end{equation}
The S-matrix is parameterized usually in the Stapp notation \cite{stapp}
\begin{equation}
%S_{lj}^{SAID} = cos \rho_{lj}^{SAID} e^{2i\delta_{lj}^{SAID}} cos
%2\epsilon_3^{SAID},
S_{lj} = \eta_{lj} e^{2i\delta_{lj}}
\end{equation}

where $\delta_{lj}$  denotes the real
part of the phase shift in the $(l,j)$th partial wave and $\eta_{lj}$ stands
for its absorptive part, the inelasticity. 
%The mixing phase between $^3D_3$ and $^3G_3$ partial waves is denoted by
%$\epsilon_3$ in the SAID nomenclature. In the current SP07 SAID solution
%$\epsilon_3$ is small and crosses zero in the region of the resonance.

For
the full partial wave amplitude in the resonating partial waves $^3D_3$ and
$^3G_3$, respectively, we take the product S-matrix approach as
used for the SAID analysis of $\pi N$ scattering \cite{SAIDpiN}: 
\begin{eqnarray}
S_{lj} = S_{lj}^B ( 1 + 2i \frac{m_R\Gamma_i}{m_R^2 - s - i
  m_R \Gamma_R} e^{2i\Phi_R}), %\nonumber
\end{eqnarray}
where $S_{lj}^B$ denotes the non-resonating background
contribution, for which we take the current SAID SP07 solution.

By doing so we assume that
\begin{itemize}
\item the energy-dependent SAID solution is not affected significantly by use
  of the data in the resonance region $T_n$ = (1.0 - 1.3) GeV. Since
  differential 
  cross section data - as we will demonstrate below - show an insignificant
  sensitivity to the resonance, the only data of relevance in this region are
  the analyzing power data at $T_n$ = 1.1 GeV. In a global SAID analysis based
  on a multitude of data such a single data set is not expected to play a
  significant role.
\item the perturbation by the resonance amplitude is small, so that no severe
  problem with unitarity arises. Multiplication of the Breit-Wigner resonance
  term with the background S-matrix in the multiplicative S-matrix
  approach helps to diminish this problem. In case of $\Phi_R$ = 0 unitarity
  is conserved by construction, otherwise one needs to check, whether for the
  resonating partial wave  $\eta \leq$~1 is still valid.
\end {itemize}

In the resonance amplitude all values are fixed with the exception of the
resonance phase $\Phi_{R}$. There are a priory no predictions for this phase
between resonance and background amplitudes. Hence it is treated as a
free parameter. 
%{\it E.g.}, in the treatment of baryon resonances in the SAID
%analysis \cite{saidpiN} of $\pi N$ scattering large resonance phases are
%obtained in most of the cases.
In the following we use the total $pn$ cross section data to fix the resonance
phase $\Phi_{R}$. 

\section{Resonance effect in $np$ scattering observables }

\begin{figure}
\centering
\includegraphics[width=0.99\columnwidth]{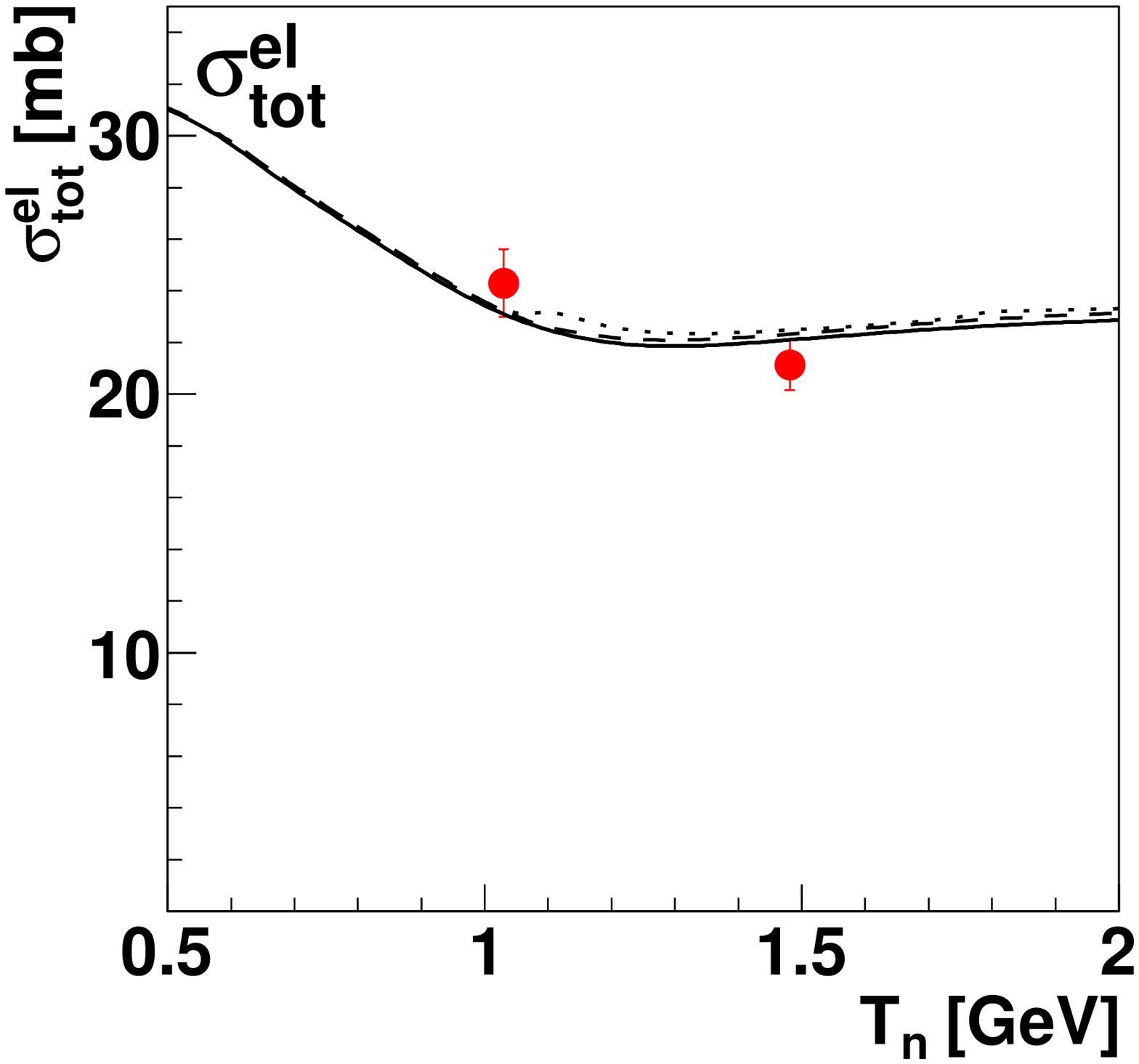}
\includegraphics[width=0.99\columnwidth]{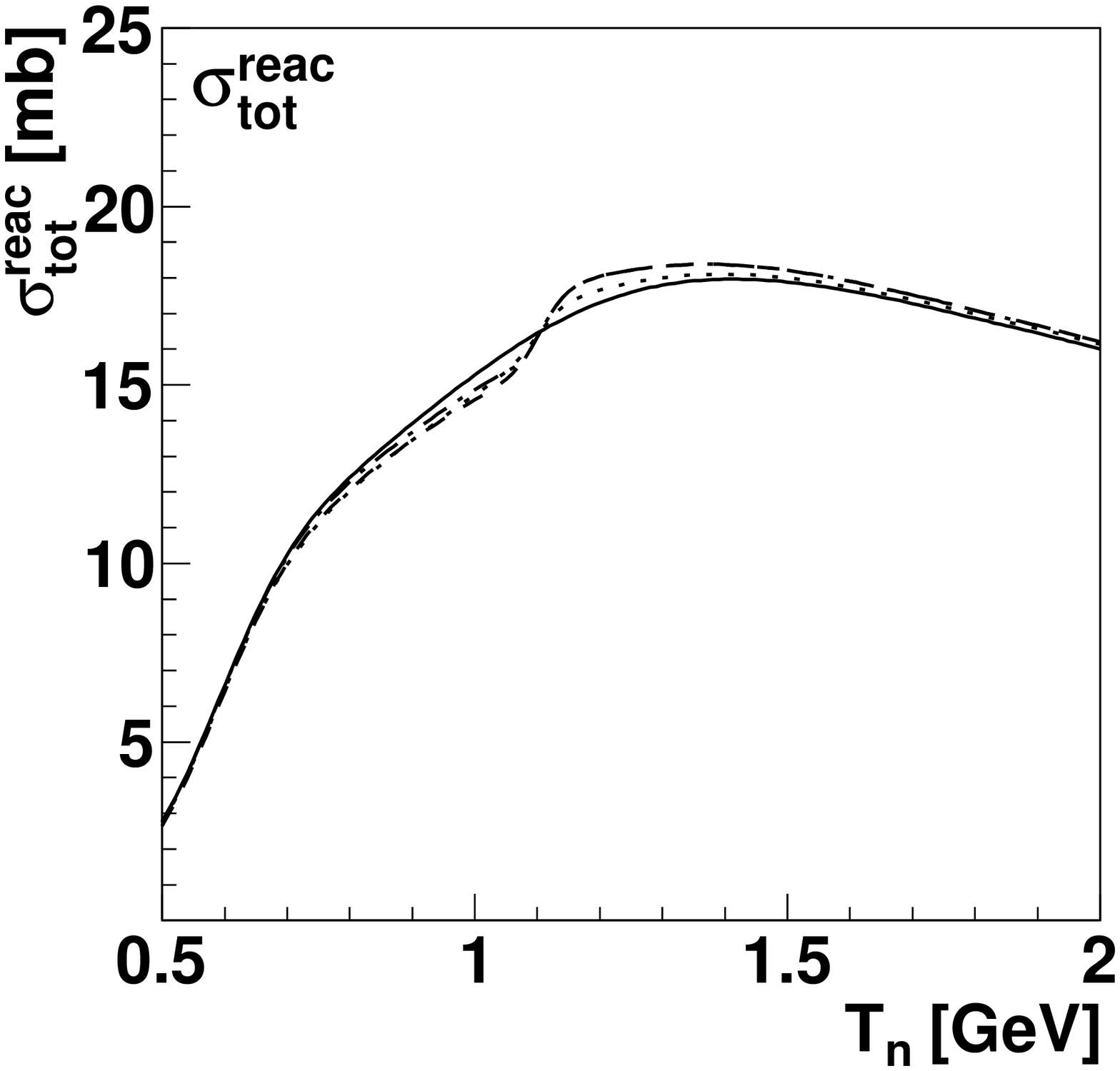}
\caption{\small Total (integral) elastic (top) and inelastic (bottom) pn cross
  sections in dependence of the incident neutron 
  energy $T_n$. The two data points are from Besliu et al. \cite{bes}. The
  solid lines denote the current SAID solution SP07 \cite{said}, the dotted
  (dashed) lines are the result, if we add the resonance amplitude in the
  $^3D_3$ ($^3G_3$) partial wave. Note that dotted and
  dashed curves lie nearly on top of each other, since the total cross
  sections are not sensitive to the partial waves' orbital angular
  momenta. The dash-dotted curves are the result, if in the $^3G_3$ case
  the inelasticity $\eta_{43}$ is constrained to unity, wherever it would
  exceed unity by adding the resonance amplitude. This concerns only the
  energy region $T_n <$ 1.1 GeV.   
}
\label{fig1}
\end{figure}

\begin{figure}
\centering
\includegraphics[width=0.99\columnwidth]{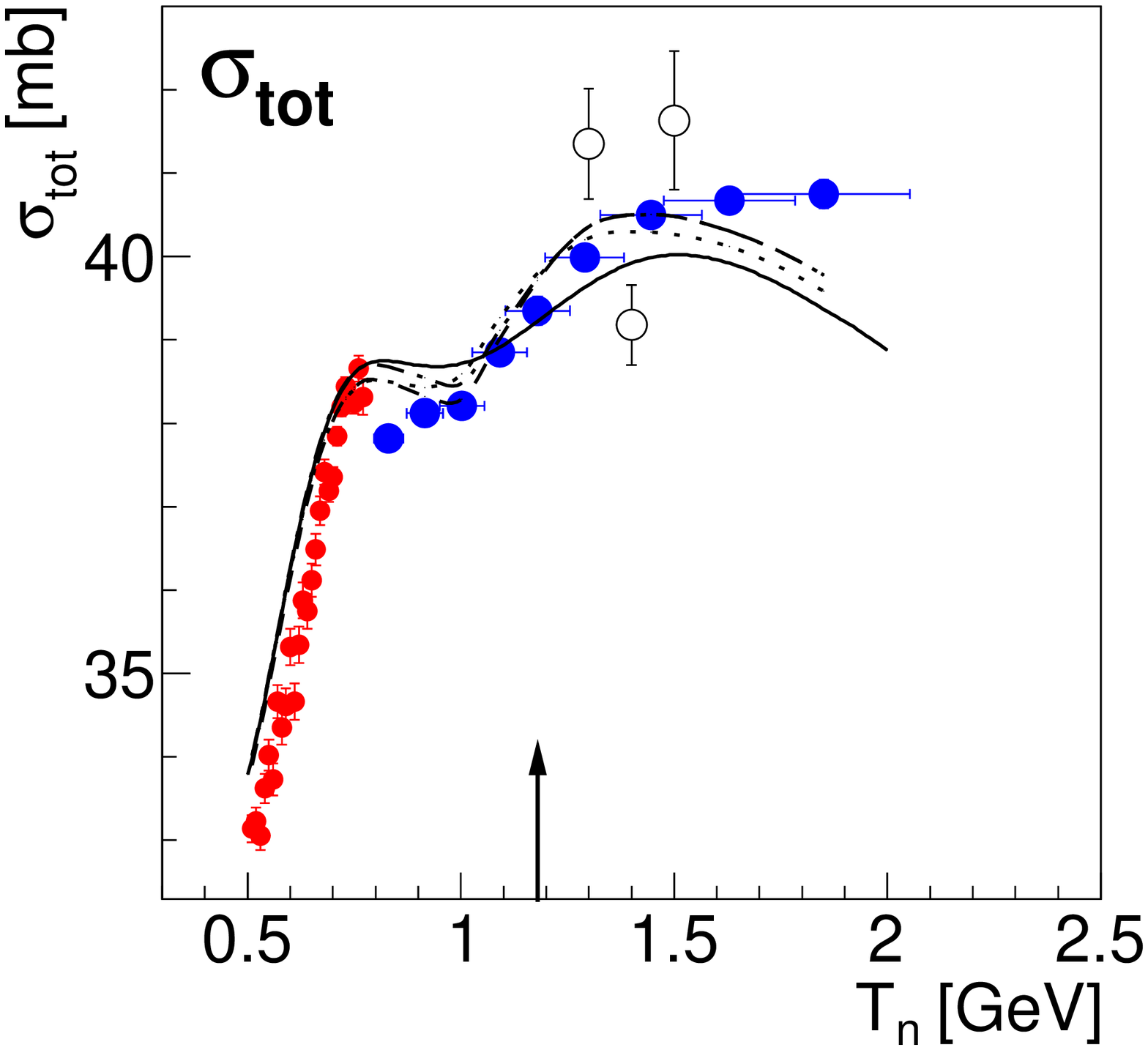}
\includegraphics[width=0.99\columnwidth]{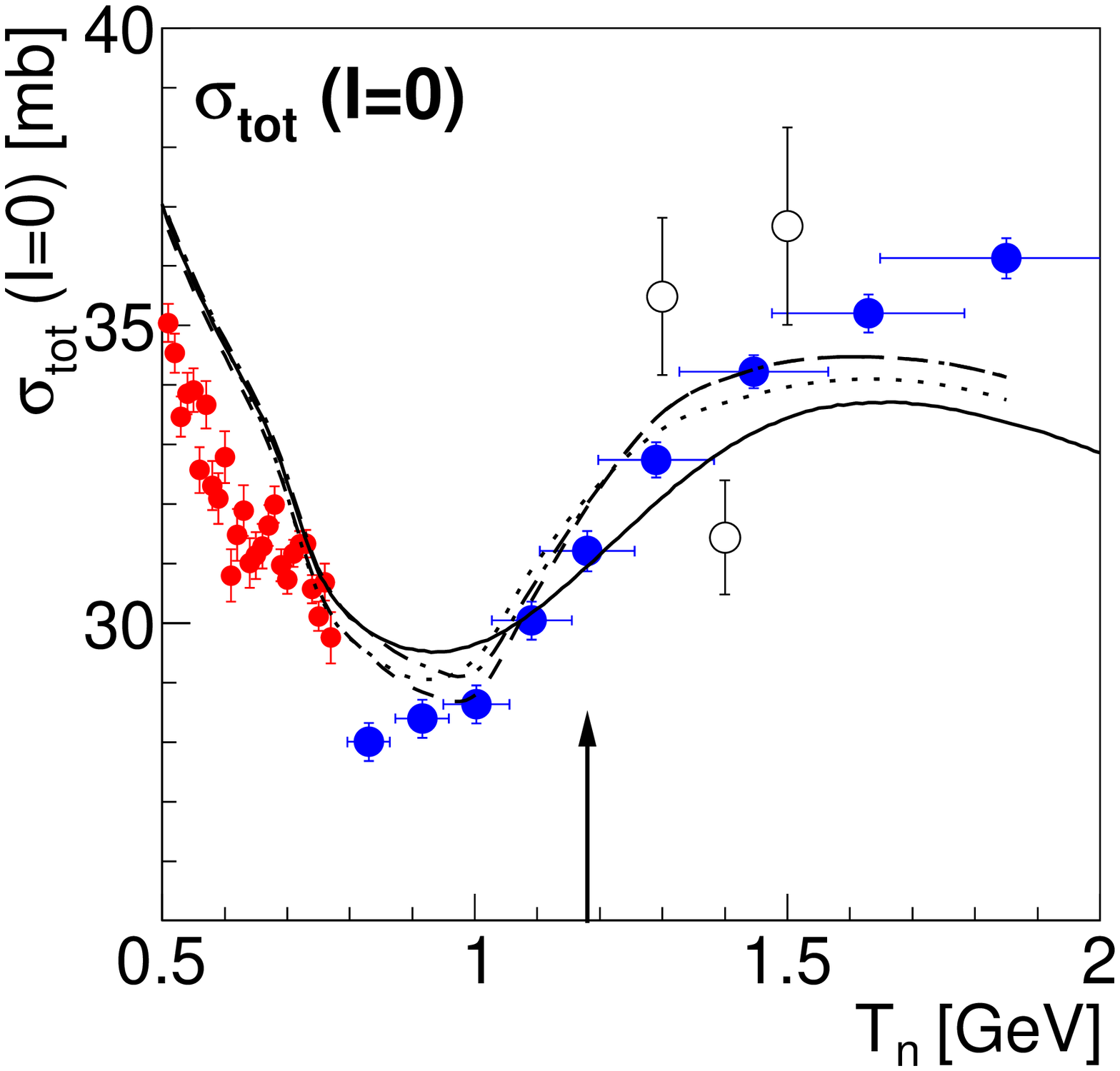}
\caption{\small Total pn cross section (top) and total isoscalar
  nucleon-nucleon cross section (bottom) in dependence of the incident neutron
  (nucleon) 
  energy $T_n$. Data (solid symbols) below 800 MeV are from Lisowski et
  al. \cite{lis} and above 800 MeV from Devlin et al. \cite{dev}. The open
  symbols represent data from Sharov et al. \cite{sha}. The  
  horizontal bars indicate the energy resolution of the incident neutrons. The
  plotted curves are averaged over these experimental energy resolutions. For
  the meaning of the curves see caption of Fig.~1. The vertical arrow
  indicates the position of the ABC resonance structure.
}
\label{fig1}
\end{figure}

The total (integral) elastic and reaction $np$ cross sections are shown in
Fig.~1. The solid curves give the current SAID solution and the dotted
(dashed) lines the result, if we add the resonance amplitude in the $^3D_3$
($^3G_3$) partial wave
with phase $ \Phi_R = -30^\circ$. As
expected from the estimate in eq. (13), the resonance effect is very small in
the integral cross sections. In addition there are no data to compare to with
the exception of two data points with large uncertainties \cite{bes}. The
experimental situation improves drastically, however, if we consider
the sum of elastic and reaction cross section, {\it i.e.}, the full total $np$
cross section, which can be accessed by 0$^\circ$ transmission measurements.

Fig. 2, top,  shows the total $np$ cross section for $T_n$~=~(0.5 - 2) GeV. The
data (solid symbols) plotted for $T_n <$ 0.8 GeV are from Lisowski et
al. \cite{lis} taken at LAMPF in a high-resolution dibaryon search. The data
plotted for $T_n >$ 0.8 GeV are from Devlin et al. \cite{dev} taken with a
neutron energy resolution of $(4 - 20)\%$ (horizontal bars in Fig. 2). Also data
from Sharov et al. \cite{sha} are shown (open symbols), which have larger
uncertainties, but are taken with a much superior neutron energy
resolution of (13 - 15) MeV. The data exhibit a pronounced jump in 
the cross section between $T_n$ = (1.0 - 1.3) GeV. This jump is remarkable,
since 
the $pp$ total cross section is completely flat in this energy region. Hence 
in the isoscalar total nucleon-nucleon cross section $\sigma_{I=0} =
2 \sigma_{pn} - \sigma_{pp}$, where the SAID values are used for
$\sigma_{pp}$, this jump appears still more pronounced (Fig. 2, bottom).   
The current SAID solution is shown by the solid lines again. Its
description of the data is only fair. In particular the observed $s$-shaped
increase in the total cross section above 1 GeV is only slightly indicated in
the SAID solution. 

If we include the resonance amplitude in the $^3D_3$ ($^3G_3$) partial wave
with a resonance phase $\Phi_R$ = 0, then we obtain a Lorentzian
shaped bump in the total cross section around $T_n \approx$ 1.1 GeV, which
roughly provides the right increase of the cross section in this energy
region, but also a fall-off thereafter, which is not
in accord with the data. To reproduce the $s$-shaped increase in the total
cross section we rather need $ \Phi_R \approx -(25 - 45)^\circ$, which
provides a  
destructive interference with the $^3D_3$ ($^3G_3$) background amplitude at
energies below the resonance mass and a constructive interference above it. This
calculation is shown in Fig. 2 by the dotted (dashed) lines. We see that the
resulting $s$-shaped pattern improves significantly the agreement with the
data. The calculations are averaged over the energy resolution of the neutron
beams (indicated by the horizontal bars in Fig.~2) used in the
experiments. This energy smearing is particularly large in the measurements of
Devlin et al. \cite{dev}. 

Putting the resonance in either 
$^3D_3$ or $^3G_3$ partial waves makes no major difference here, since the
total cross sections are not sensitive to the partial waves' orbital angular
momenta. Slight differences arise from the fact that we have different
momentum dependences for $^3D_3$ and $^3G_3$ partial waves --- see eq. (15) --
and in particular from the fact that the resonance amplitude is multiplied by
the background amplitude -- see eq.~(22), where the real parts of $^3D_3$ and
$^3G_3$ phase shifts differ by more than 10$^\circ$.

\begin{figure}
\includegraphics[width=0.49\columnwidth]{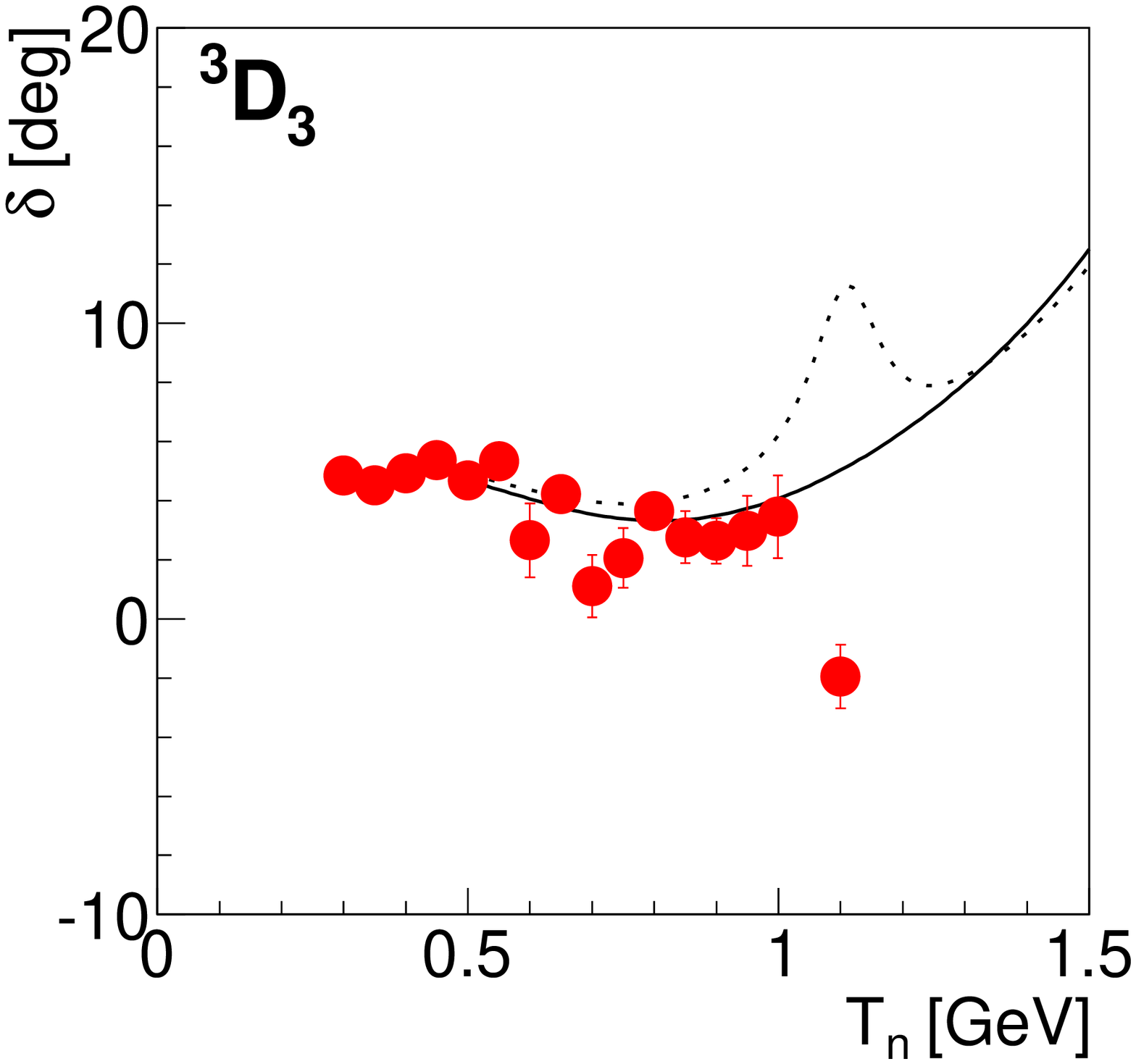}
\includegraphics[width=0.49\columnwidth]{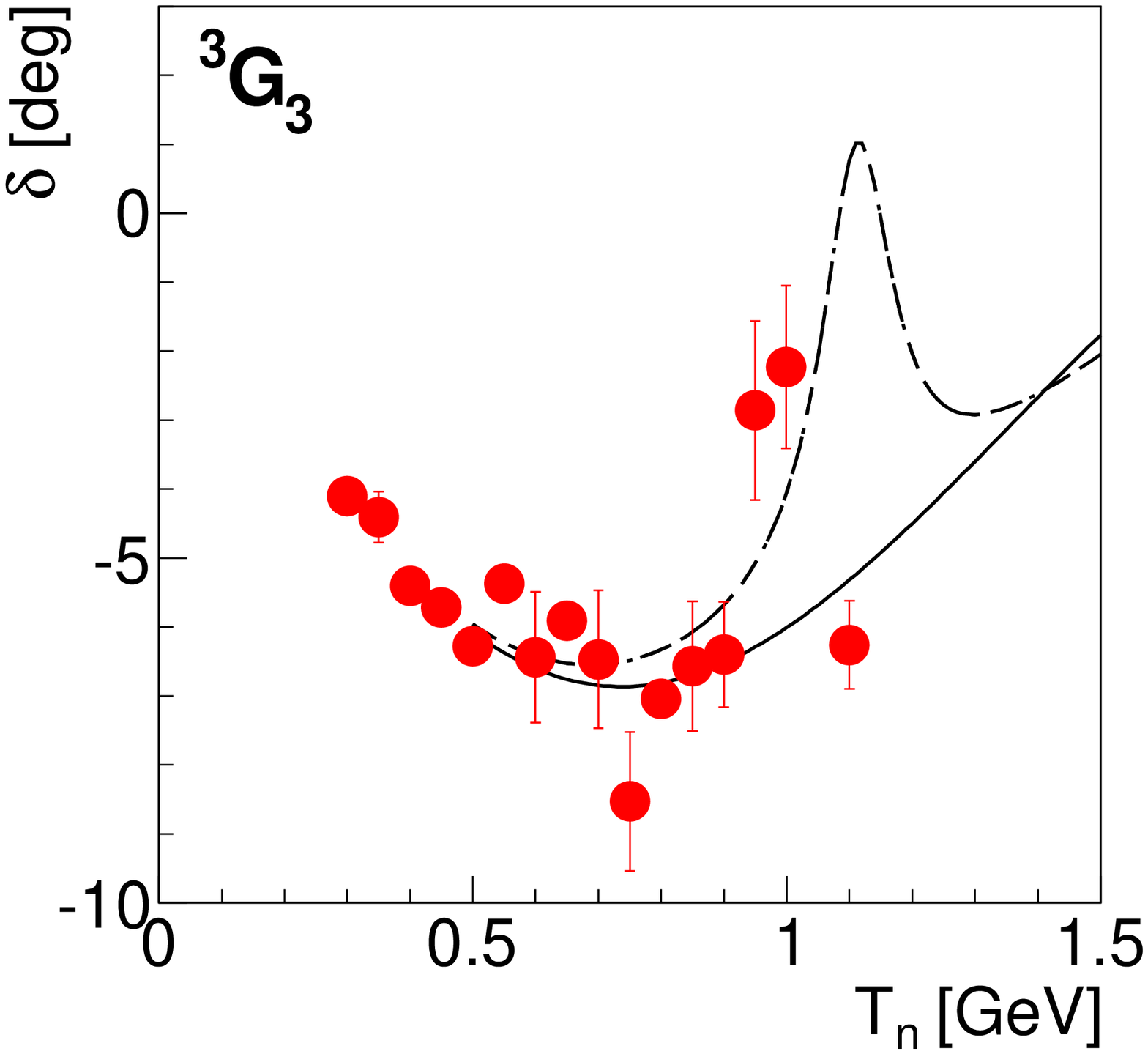}
\includegraphics[width=0.49\columnwidth]{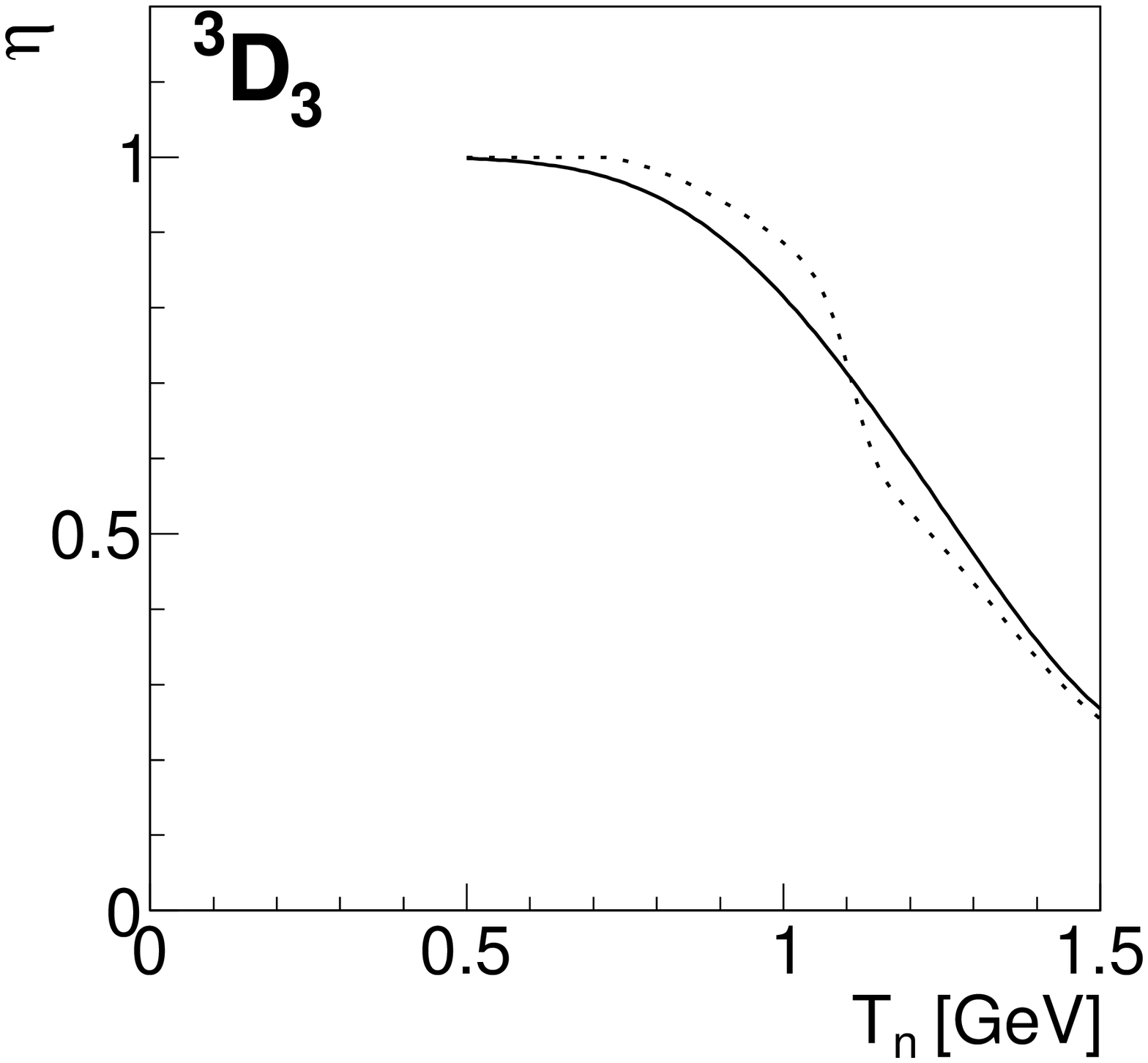}
\includegraphics[width=0.49\columnwidth]{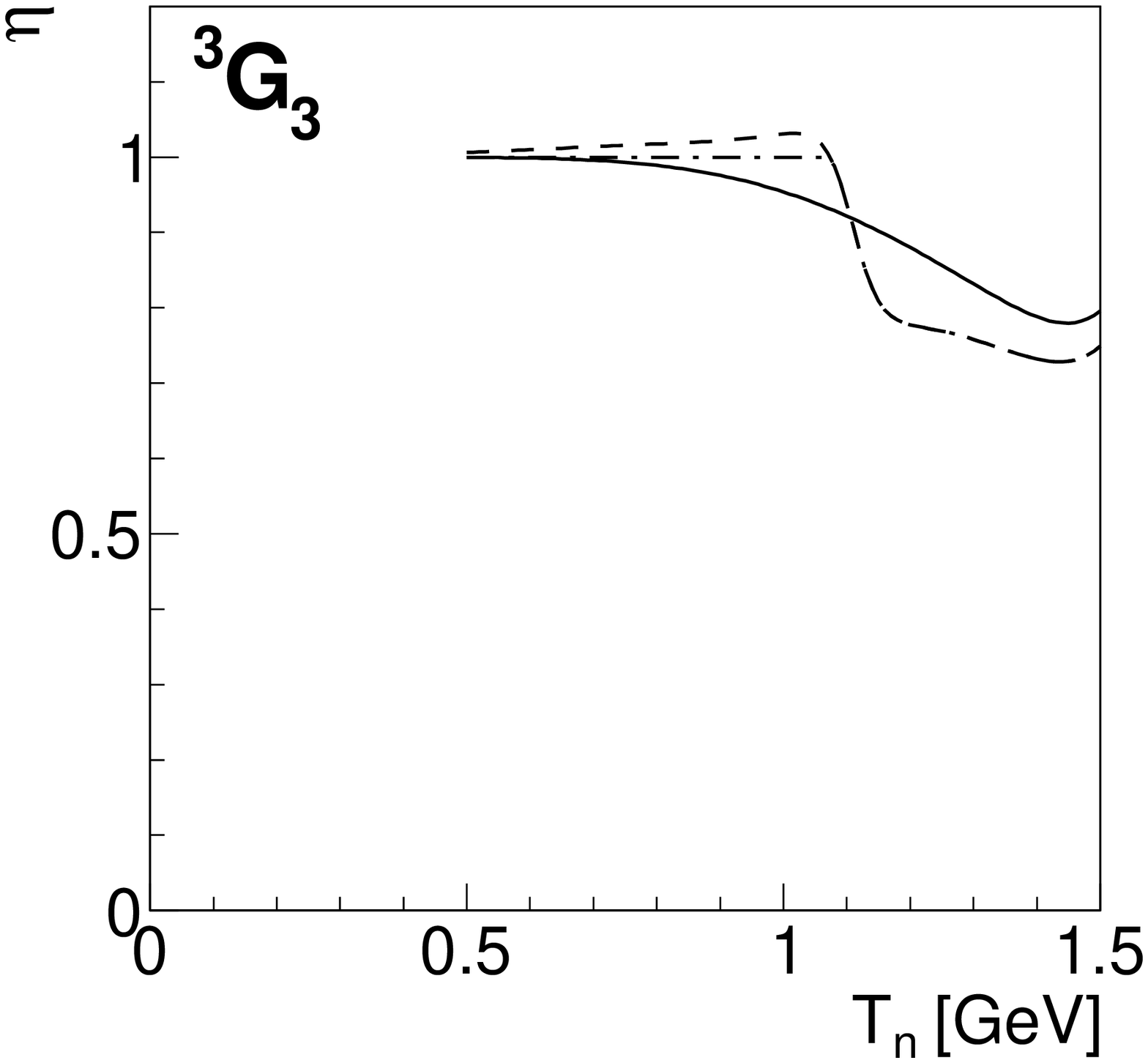}
\includegraphics[width=0.49\columnwidth]{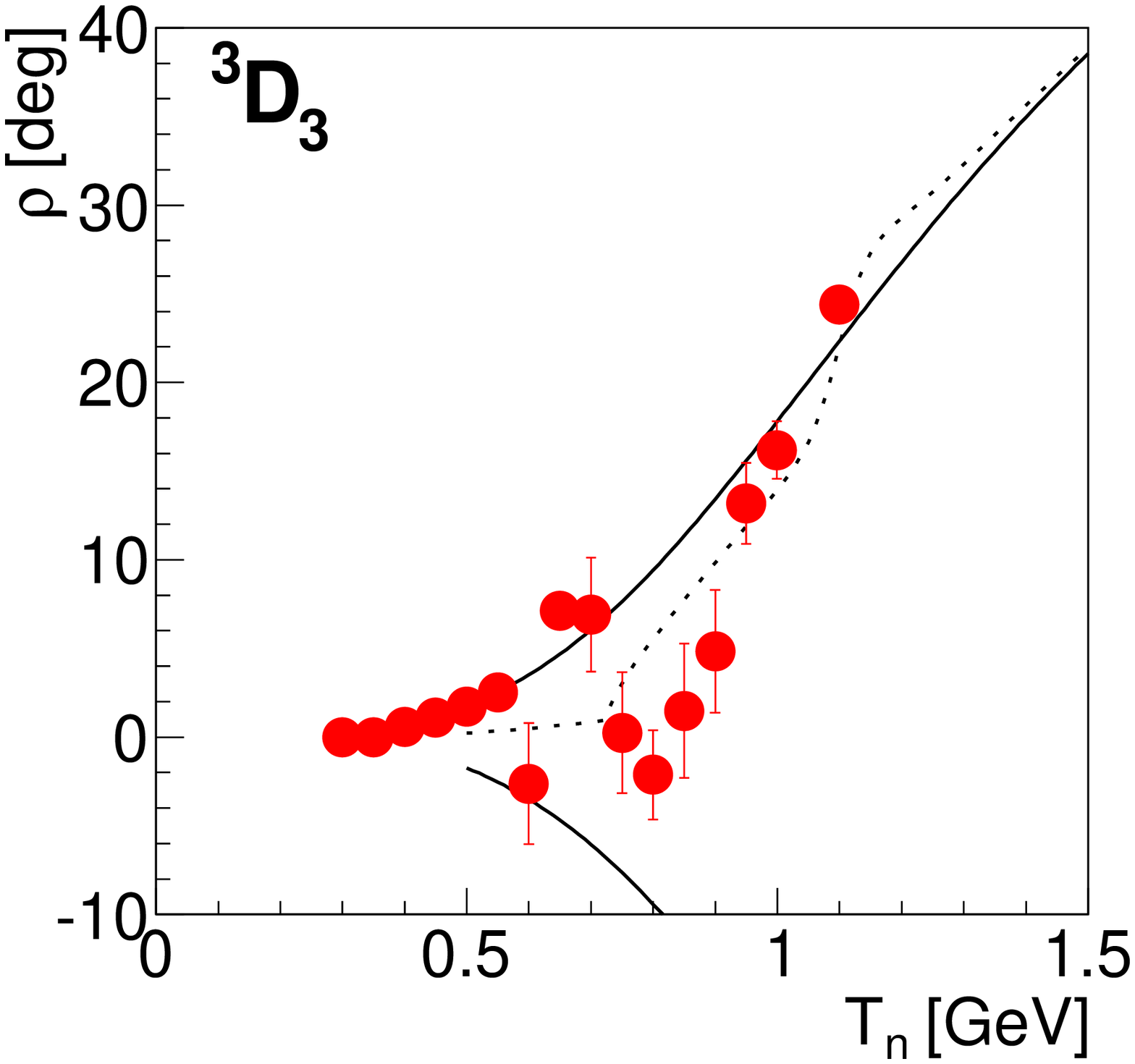}
\includegraphics[width=0.49\columnwidth]{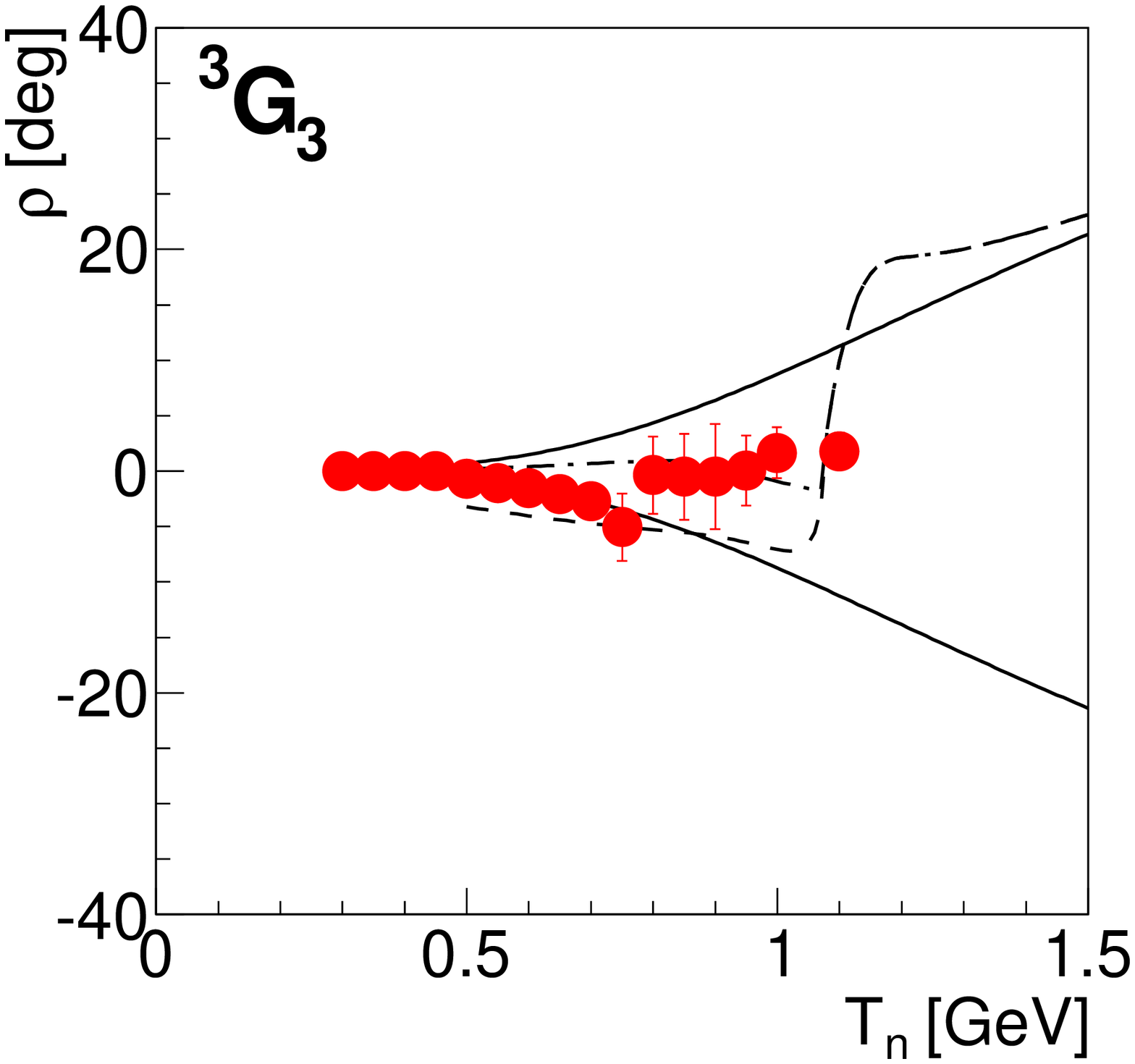}
\caption{\small Energy dependence of the phase shifts for $^3D_3$ (left) and
  $^3G_3$ (right) partial waves. The real parts $\delta_{lj}$ are shown at the
  top, the imaginary parts below either as inelasticity $\eta_{lj}$ (Stapp
  notation \cite{stapp}) in the middle or as $\rho_{lj}$ phase in the SAID
  convention \cite{said,arndtroper} at the bottom. The 
  solid lines and symbols denote the SAID SP07 energy dependent and single
  energy solutions, respectively \cite{said}. 
%Note that since $cos(-\rho) = cos \rho$, the sign of $\rho$ is ambiguous. 
  Since the sign of $\rho$ does not enter \cite{arndtroper}, we plot the SAID
  solution for $\rho$ for both signs. Dotted and dashed curves show the
  results of including the resonance amplitude in $^3D_3$ and $^3G_3$ partial
  waves, respectively. The dash-dotted curve results, if in the $^3G_3$ case
  the inelasticity $\eta_{34}$ is constrained to unity, wherever it would
  exceed unity by adding the resonance amplitude. This concerns only the
  energy region $T_n <$~1.1~GeV. 
}
\label{fig1}
\end{figure}

The pase shifts for  $^3D_3$ and $^3G_3$ partial waves in the energy region of
interest are depicted in Fig.~3.
For the $^3D_3$-case the inclusion of the resonance with $\Phi_R \approx
-(25 - 
45)^\circ$ does not cause problems with unitarity, since the background
inelasticity $\eta_{23}^B$ is already much below unity in the energy
region of the resonance.  

For the $^3G_3$-case the situation is much more delicate, since
$\eta_{43}^B$ is still close to unity in the resonance region -- with the
consequence that the the total $\eta_{43}$ gets slightly above unity for
energies below 1.1 GeV. This points to the necessity that the background
amplitudes would need to be readjusted, when taking into account the resonance
explicitly. Since this would mean a major effort much beyond the
scope of  this work, where the main emphasis is to demonstrate the basic 
effect of the resonance on the observables, we  
demand for simplicity $\eta_{43} = 1$ in the region, where it would exceed
unity.
(Effectively, this means that we readjust the background inelasticity
$\eta_{43}^B$ accordingly.)
This constrained calculation is shown in the figures by the dash-dotted
lines. As expected, the calculation for the total cross sections 
falls now speedily back to the SAID solution in energy region below 1.1 GeV,
where $\eta_{43}$ is now constrained to unity. As we will show below in 
Fig.~5, this constraint has only tiny effects on the differential
$np$-scattering observables at energies below 1.1 GeV.

After having succeeded in improving the description of the total cross section 
data substantially by inclusion of the resonance amplitude in $^3D_3$ or
$^3G_3$ partial 
waves, we consider now the resonance effect in the differential observables.
In contrast to the situation for the integral cross section,  it will make here
a substantial difference, whether the resonance is in the $^3D_3$ or the  
$^3G_3$ partial wave due to the different angular dependences of these
partial waves -- in particular in the analyzing power $A_y$, as we will
demonstrate in the following.

\begin{figure} [t]
\centering
\includegraphics[width=0.49\columnwidth]{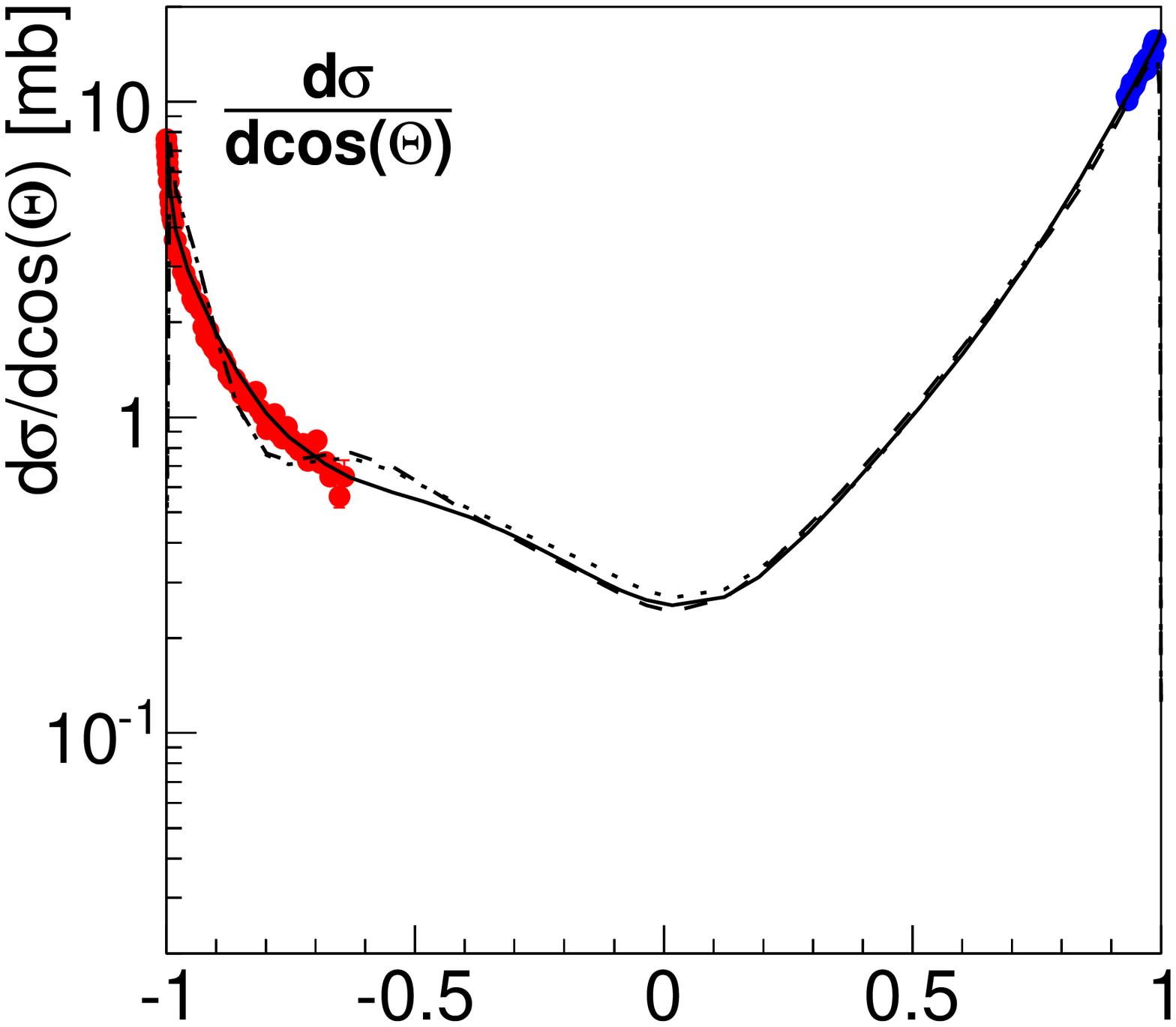}
\includegraphics[width=0.49\columnwidth]{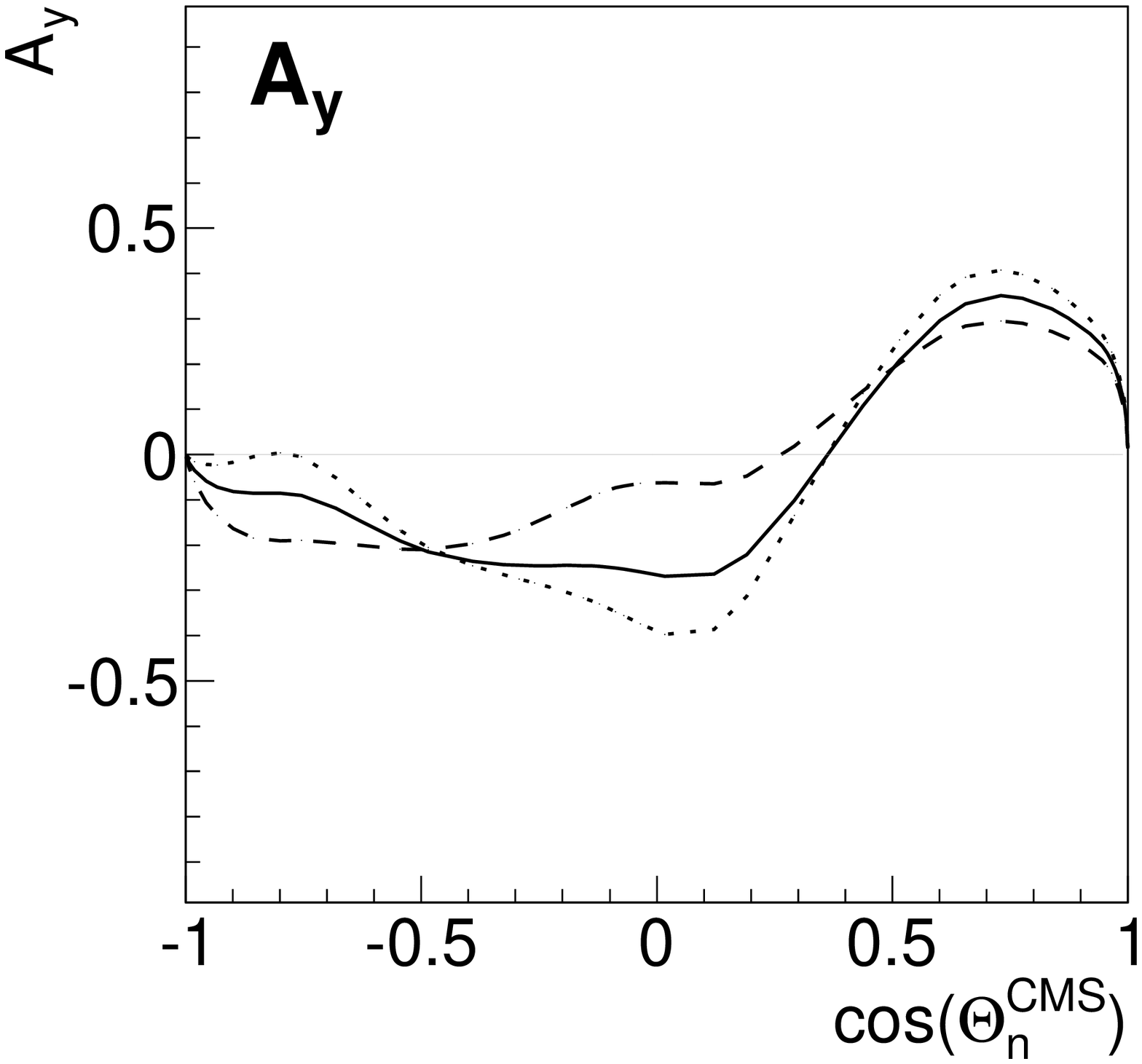}
\includegraphics[width=0.49\columnwidth]{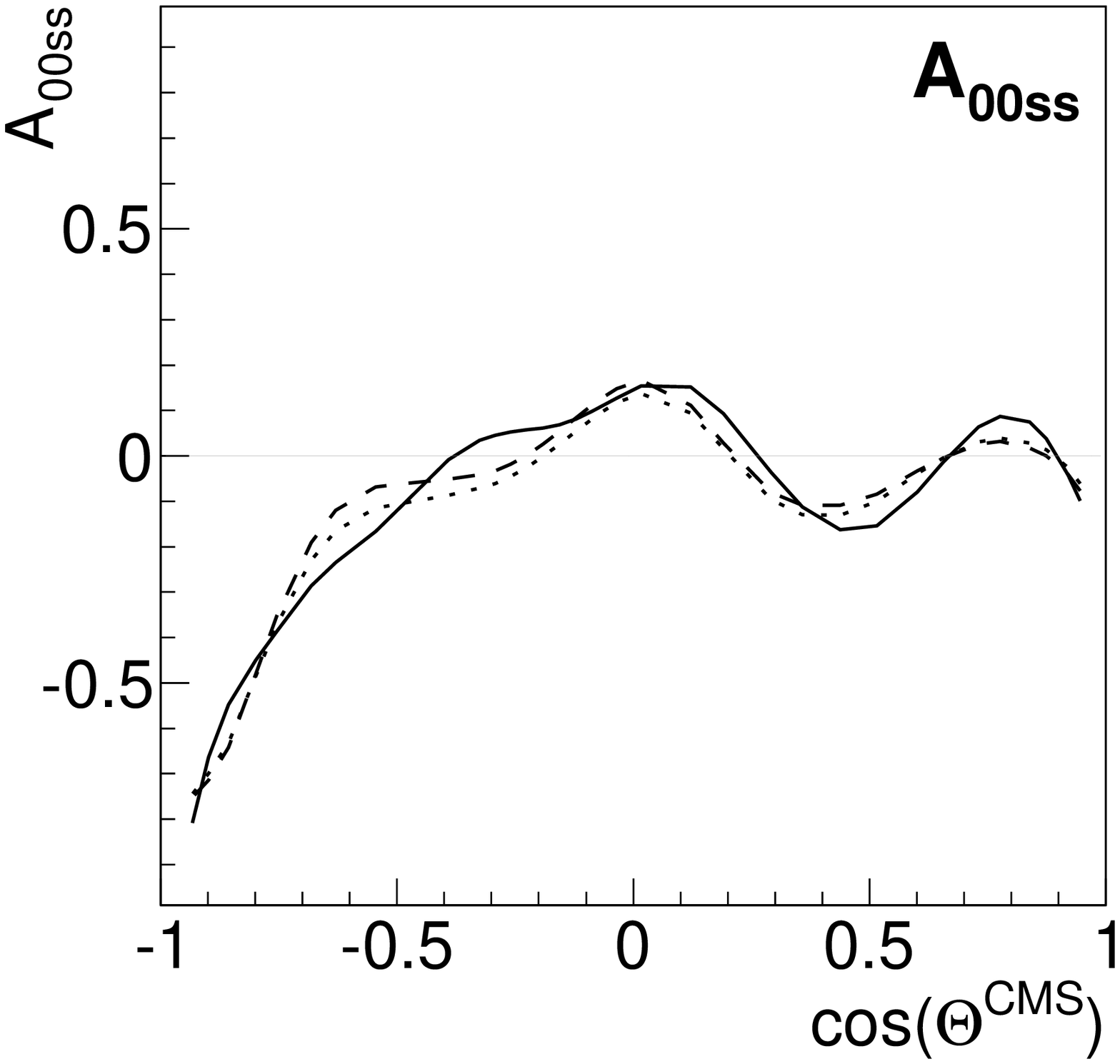}
\includegraphics[width=0.49\columnwidth]{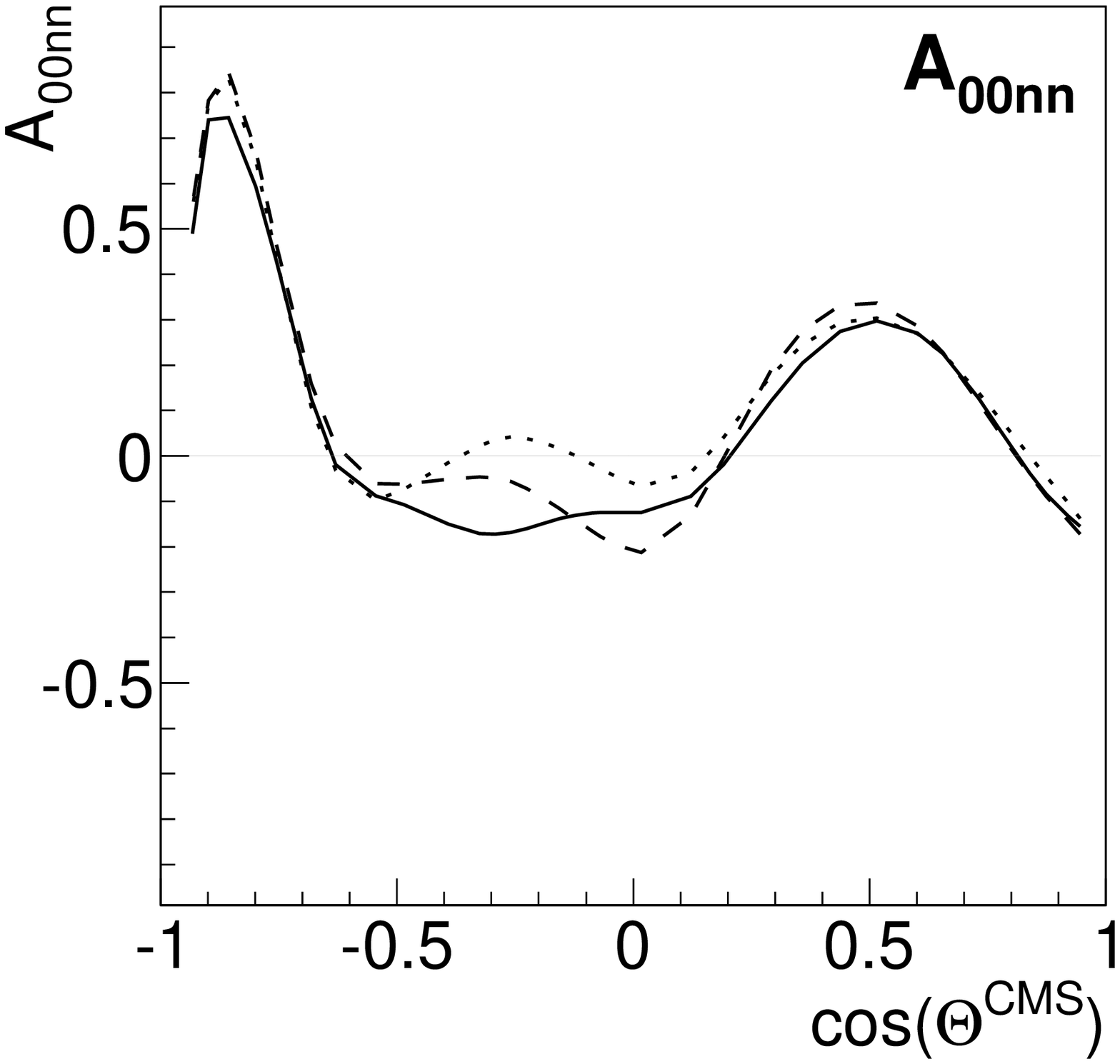}
\includegraphics[width=0.49\columnwidth]{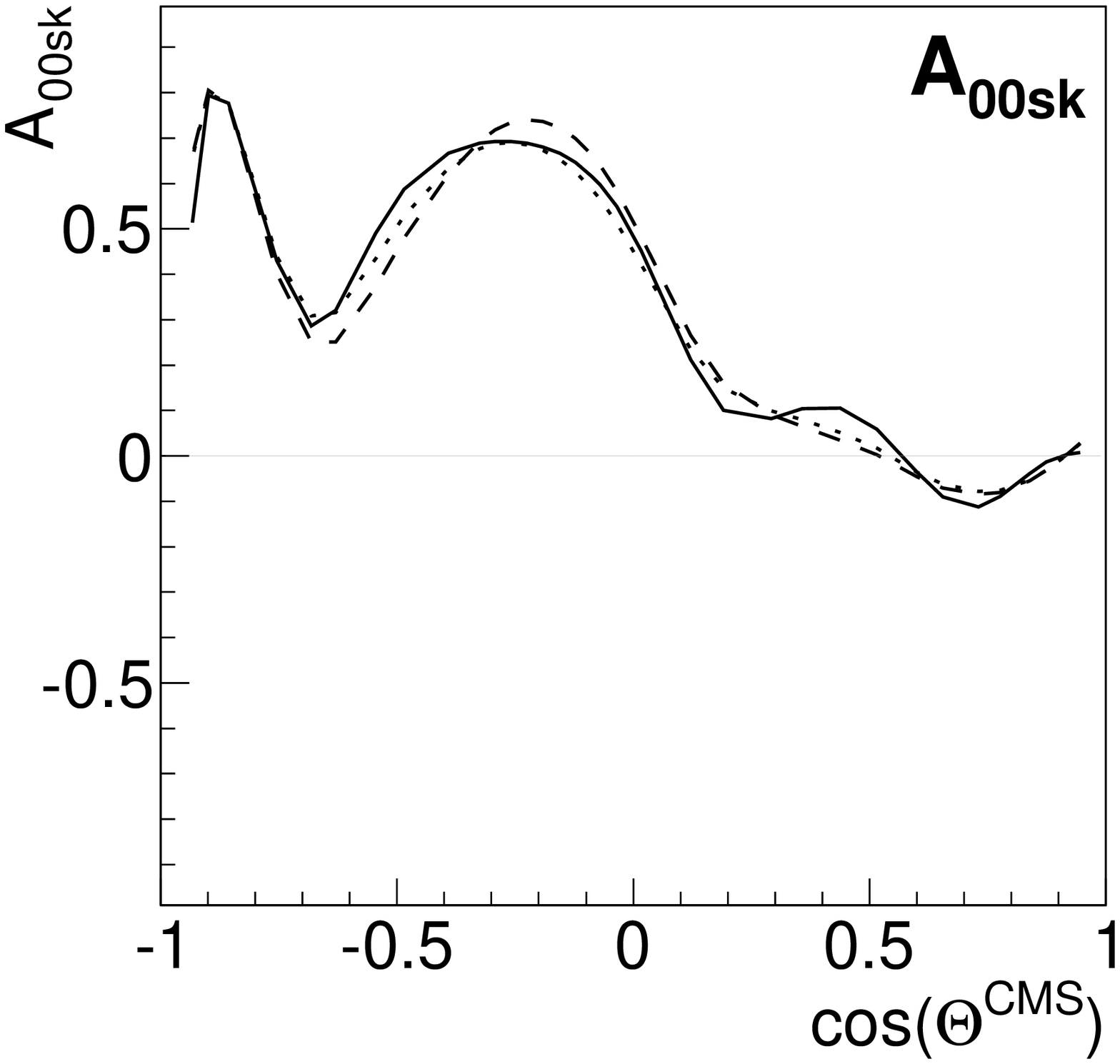}
\includegraphics[width=0.49\columnwidth]{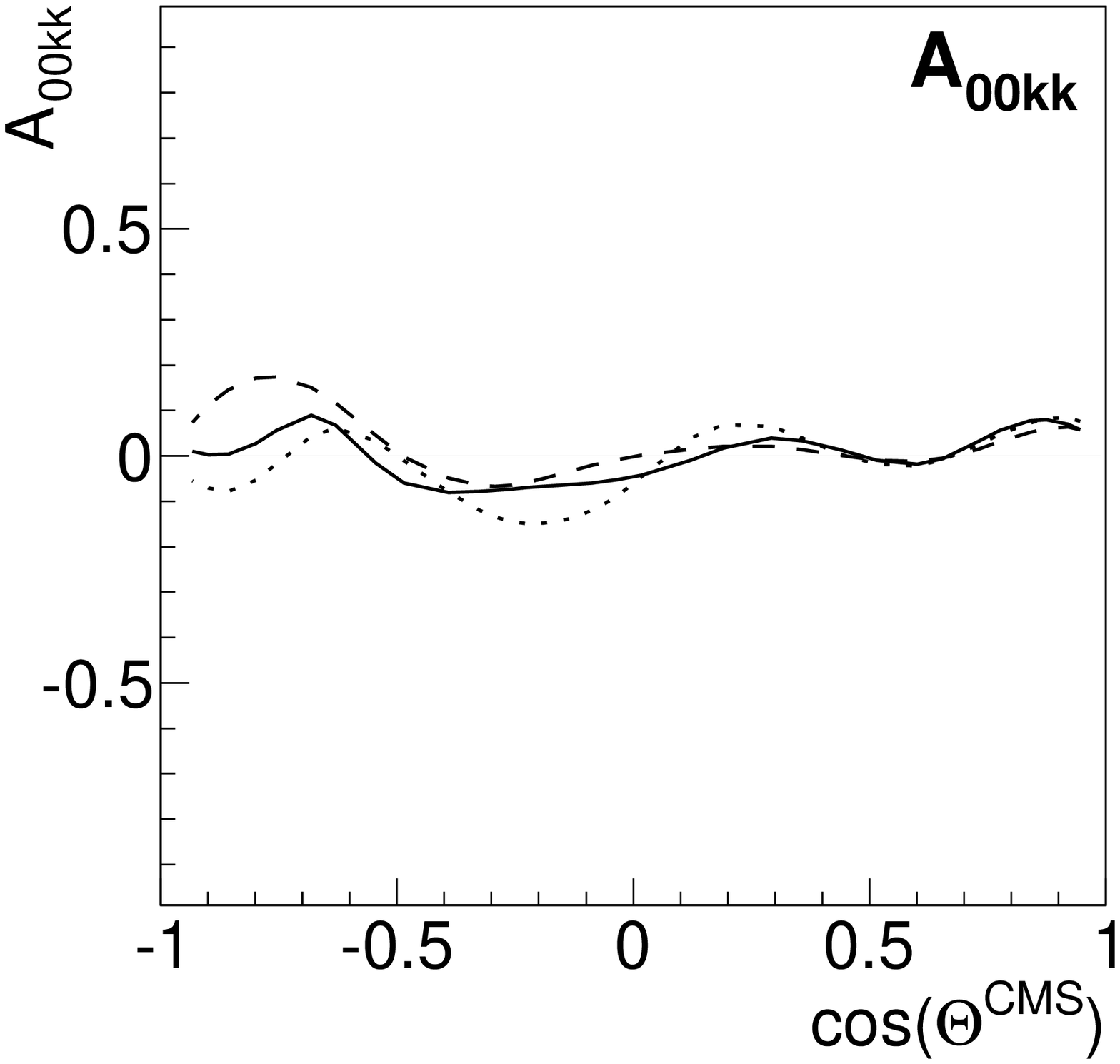}
\caption{\small Differential distributions of cross section $d\sigma /
  dcos(\Theta)$, vector analyzing power $A_y$ and spin correlation coefficients
  $A_{00ij}$ at $T_n$ = 1.13 GeV corresponding to the resonance energy $\sqrt
  s$ = 2.37 GeV. For the meaning of the curves see caption of Fig.~1. For the
  differential cross section data are plotted for the nearby energies 
  $T_n$ = 1.118 GeV \cite{bizard} and $T_n$ = 1.135 GeV \cite{terrien}.
}
\label{fig1}
\end{figure}

\begin{figure}
\centering
\includegraphics[width=0.99\columnwidth]{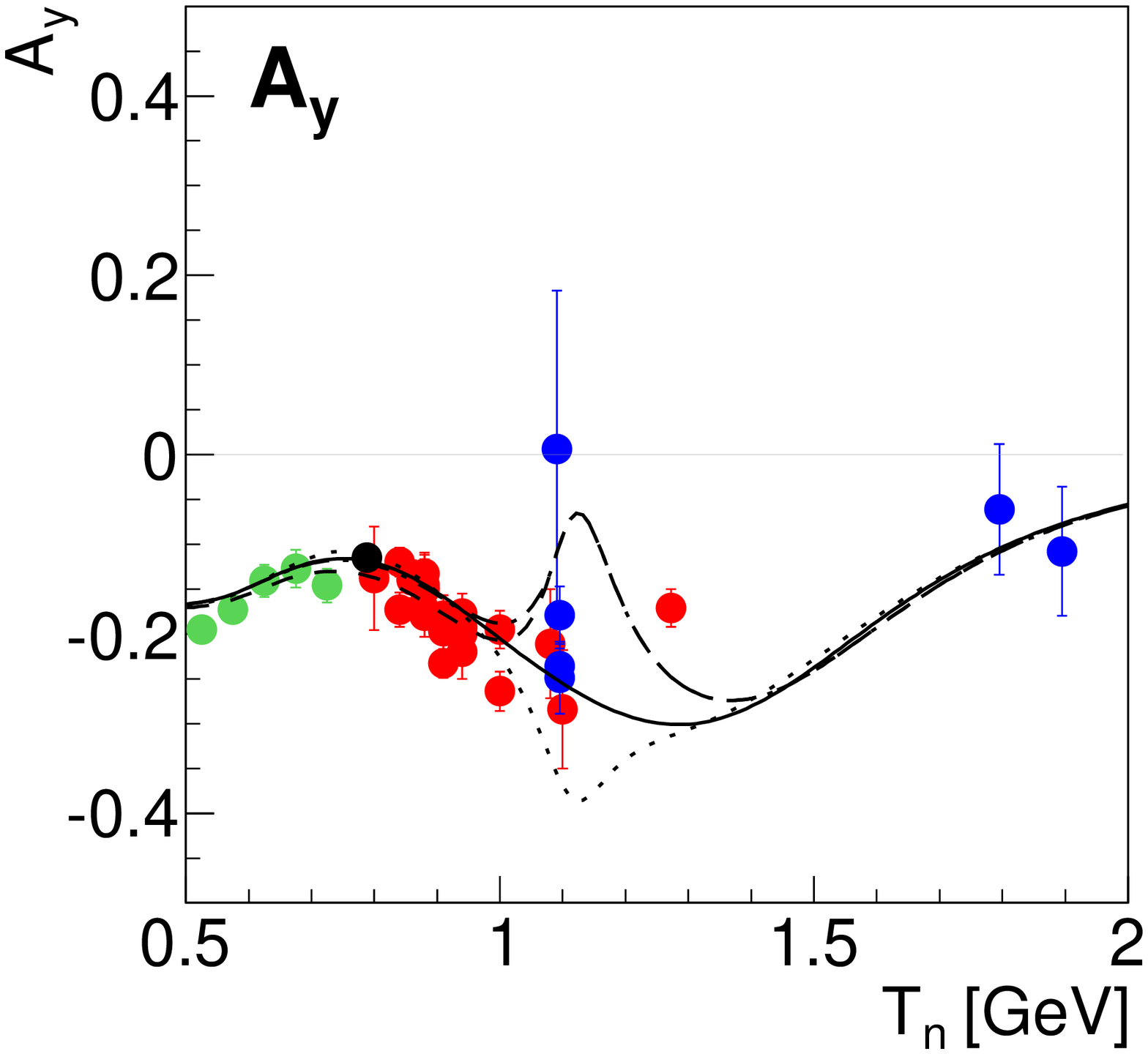}
\caption{\small Energy dependence of the vector
  analyzing power at $\Theta_{cm}$ = 83$^\circ$. The plotted data are from
  \cite{ball,les,new,arn,bal1,mcn,sak,gla,mak}. For the meaning of the curves
  see Fig.~1.
}
\label{fig1}
\end{figure}

Fig. 4 shows the angular distributions of differential cross section $d\sigma /
  dcos(\Theta)$, vector analyzing analyzing power $A_y$ and spin correlation
  coefficients $A_{00ij}$ at $T_n$ = 1.13 GeV corresponding to
the resonance energy $\sqrt s$ = 2.37 GeV, where we expect the effect of the
resonance on the observables to be largest. At this energy there are only data
for the differential cross section at small scattering angles. The solid lines
denote the current SAID 
solution, the dotted (dashed) lines give the result with the resonance
amplitude added in the $^3D_3$ ($^3G_3$) partial wave. As expected from the
discussion of the integral elastic cross section the
resonance effect is tiny  in the differential cross
section, however, sizably in the polarization observables. It is
largest in the analyzing power $A_y$, which solely depends on interference
terms.  The resonance effects are particularly notable at intermediate
angles, where the differential cross section gets smallest. We also see that
$^3D_3$ and $^3G_3$ resonance contributions lead to opposite
effects there. This provides the opportunity to disentangle these
contributions by $A_y$ measurements.

The decomposition of the $np$-scattering observables into partial wave
amplitudes is given in Ref. \cite{arndtroper}. Accordingly we have for the
analyzing power:

\begin {equation}
 d\sigma / dcos(\Theta) * A_y \sim Im{(H_3 + H_5) H_4^*}
\end {equation}

with $H_i$ containing sums over partial wave amplitudes with total angular
momenta $j_0 = j = L$, $j_- = L - 1$ and $j_+ = L + 1$. $H_3$ contains terms
being proportional either to the Legendre polynomials $P_j$ or to the associated
ones $P_j^1$. In $H_5$ there are terms only proportional to $P_j$ and in $H_4$
only proportional to $P_j^1$. In particular, the structure of $H_4$ for $j =
3$ is as follows:
\begin {equation}
H_4(j = 3) \sim [4(T_{L=4} - 3 T_{L=2}) + \sqrt{12} T_{L=3}] P_3^1,  
\end {equation}
where the T-matrix elements contain the complex phase shifts. We see that a
resonance effect in $^3D_3$ and $^3G_3$ enters with opposite sign and is
proportional to $P_3^1$ in both cases. Hence the resonance effect vanishes at
the zeros of $P_3^1$, which is the case at $cos(\Theta) = \pm 1 / \sqrt 5 = \pm
0.447$ corresponding to $\Theta = 63.4^\circ$ and 116.6$^\circ$. At these
angles the predictions with and without resonance in $^3D_3$ 
or $^3G_3$ cross each other -- see Fig.~4, top right. $P_3^1$ is maximal
at $cos(\Theta) = \pm \sqrt {11 / 15} = \pm 0.856$ and minimal at
$cos(\Theta)~=~0$. Since at the latter the differential cross section is
minimal and much 
lower than at  $cos(\Theta) = \pm 0.856$ -- see Fig. 3, left --, the resonance
effect in $A_y$ gets maximal at $cos(\Theta) = 0$, {\it i.e.} at $\Theta =
90^\circ$.

In Fig. 5 we plot the energy dependence of $A_y$ near $\Theta = 90^\circ$, 
the angular region, where we find the largest resonance effects and where also a
large amount of data are available, in particular from neutron-proton
scattering experiments at Saclay \cite{ball,les}. Since the angular dependence
around $\Theta = 90^\circ$ is small, we plot in Fig.~5 the energy
dependence at $\Theta = 83^\circ$, where the situation of available data
\cite{ball,les,new,arn,bal1,mcn,sak,gla,mak} is more favorable than at $\Theta
= 90^\circ$. The meaning of the drawn curves is the same as in Fig.~4. A
significant resonance effect shows up within the energy region $T_n$ = (1.0 -
1.3) GeV. The effect is opposite in sign for the resonance residing in $^3D_3$
or $^3G_3$ partial waves. 
Note also that the calculations with (dashed) and without (dash-dotted) the
constraint $\eta_{43} \leq$ 1 exhibit only small differences for energies below
1.1 GeV -- well within uncertainties of currently available data. This is not
unexpected, 
since according to eqs. (20) - (24) the analyzing power is mainly sensitive to
the real part of the phase shift.

\section{Conclusions}

Summarizing, we have shown that the  $I(J^p) = 0(3^+)$ resonance structure
found in the basic double-pionic fusion process $pn \to d\pi^0\pi^0$ is
consistent with existing $np$ scattering data. The effect of such a $s$-channel
resonance is significant in specific $np$ observables. In particular it
improves considerably the description of the total cross section  
beyond 1 GeV. Among the differential observables the vector analyzing power 
exhibits the largest sensitivity to the resonance. However, for a crucial test
of the resonance hypothesis and a meaningful separation of $^3D_3$ and $^3G_3$
resonance contributions high-precision data are needed for the energy region
$T_n$ = (1.0 - 1.3) GeV. Such measurements have actually been carried
out very recently with the WASA detector at COSY and the data analysis has
started. The WASA detector installed at the COSY ring is particularly suited 
for analyzing power measurements in the intermediate angle region, which -- as
we have demonstrated here -- is of main interest for the search of resonance
effects in $np$ scattering.

We finally note that on the issue of the $I(J^p) = 0(3^+)$ resonance
structure meanwhile a first three-body Faddeev calculation with full
relativistic kinematics and based on hadron dynamics has been carried out by
Gal and Garzilaco \cite{gal}. They find, indeed, a resonance with just these
quantum numbers at a mass of 2.36(2) GeV in agreement with the experimental
observation.  
 
\section{Acknowledgments}
We acknowledge valuable discussions on this matter with J. Haidenbauer,
C. Hanhart, F. Hinterberger, A. Kacharava, I. Strakovsky, H. Str\"oher,
G.J. Wagner, C. Wilkin, A. Wirzba and R. Workman. 
We are indebted to C. Elster for using her partial wave code by one of us
(A.P.).  
This work has been
supported by the BMBF (06TU9193) and the Forschungszentrum J\"ulich (COSY-FFE).

\end{document}